\documentclass[aps, 10pt, prl,notitlepage, twocolumn, superscriptaddress,nofootinbib]{revtex4-1}

\usepackage{amsmath}
\usepackage{amssymb}
\usepackage{amsfonts}
\usepackage[utf8]{inputenc}
\usepackage[T1]{fontenc}
\usepackage{mathrsfs}
\usepackage{anyfontsize}

\usepackage[usenames,dvipsnames]{xcolor}

\usepackage{newtxtext}
\usepackage{newtxmath}
\usepackage{tensor}
\usepackage{bm}

\usepackage{graphicx}
\usepackage{epsfig}
\usepackage{epstopdf}

\usepackage{natbib}

\usepackage[normalem]{ulem}

\def\jnl@style{\it}
\def\aaref@jnl#1{{\jnl@style#1}}

\def\aaref@jnl#1{{\jnl@style#1}}

\def\aj{\aaref@jnl{AJ}}                   
\def\apj{\aaref@jnl{ApJ}}                 
\def\apjl{\aaref@jnl{ApJ}}                
\def\apjs{\aaref@jnl{ApJS}}               
\def\apss{\aaref@jnl{Ap\&SS}}             
\def\aap{\aaref@jnl{A\&A}}                
\def\aapr{\aaref@jnl{A\&A~Rev.}}          
\def\aaps{\aaref@jnl{A\&AS}}              
\def\mnras{\aaref@jnl{Mon.~Not.~Roy.~Astron.~Soc.}}             
\def\prd{\aaref@jnl{Phys.~Rev.~D}}        
\def\prc{\aaref@jnl{Phys.~Rev.~C}}  
\def\prl{\aaref@jnl{Phys.~Rev.~Lett.}}    
\def\qjras{\aaref@jnl{QJRAS}}             
\def\skytel{\aaref@jnl{S\&T}}             
\def\ssr{\aaref@jnl{Space~Sci.~Rev.}}     
\def\zap{\aaref@jnl{ZAp}}                 
\def\nat{\aaref@jnl{Nature}}              
\def\aplett{\aaref@jnl{Astrophys.~Lett.}} 
\def\apspr{\aaref@jnl{Astrophys.~Space~Phys.~Res.}} 
\def\physrep{\aaref@jnl{Phys.~Rep.}}      
\def\physscr{\aaref@jnl{Phys.~Scr}}       
\def\commat{\aaref@jnl{Comm.~Math.~Phys.}}              
\def\science{\aaref@jnl{Science}}               
\def\cqg{\aaref@jnl{Classical Quant.~Grav.}}            
\def\jpcs{\aaref@jnl{JPCS}}                                     
\def\ijmpd{\aaref@jnl{Int.~J.~Mod.~Phys.~D}}                    
\def\grg{\aaref@jnl{Gen.~Relat.~Gravit.}}               
\def\rpp{\aaref@jnl{Rep.~Prog.~Phys.}}          
\def\npa{\aaref@jnl{Nucl.~Phys.~A}}        
\def\lrr{\aaref@jnl{Living Rev.~Rel.}}                   
\def\jcap{\aaref@jnl{J.~Cosmology Astropart.~Phys.}}    
\def\rmp{\aaref@jnl{Rev.~Mod.~Phys.}}   

\allowdisplaybreaks[1]

\begin{document}

\title{Topological neutron stars in tensor-multi-scalar theories of gravity }

\author{Daniela D. Doneva}
\email{daniela.doneva@uni-tuebingen.de}
\affiliation{Theoretical Astrophysics, Eberhard Karls University of T\"ubingen, T\"ubingen 72076, Germany}
\affiliation{INRNE - Bulgarian Academy of Sciences, 1784  Sofia, Bulgaria}

\author{Stoytcho S. Yazadjiev}
\email{yazad@phys.uni-sofia.bg}
\affiliation{Theoretical Astrophysics, Eberhard Karls University of T\"ubingen, T\"ubingen 72076, Germany}
\affiliation{Department of Theoretical Physics, Faculty of Physics, Sofia University, Sofia 1164, Bulgaria}
\affiliation{Institute of Mathematics and Informatics, 	Bulgarian Academy of Sciences, 	Acad. G. Bonchev St. 8, Sofia 1113, Bulgaria}


\begin{abstract}
In the present paper we demonstrate that there exists a new type of neutron stars in the tensor-multi-scalar theories of gravity. 
We call this new type of neutron stars {\it topological neutron stars}. In addition to the standard characteristics of the usual neutron stars
the topological neutron stars are also characterized by a topological charge. We numerical construct explicit examples of topological neutron stars in tensor-multi-scalar theories whose target space is $\mathbb{S}^3$. Besides the topological charge the topological neutron stars also exhibit other attractive features which can place them  among the realistic compact objects that could exist in Nature. For example they possess zero scalar charge and thus  evades the strong binary pulsar constraints on the dipole scalar radiation.            
 
\end{abstract}


\maketitle

\section{Introduction}

In the era of gravitational wave astronomy it is expected that gravitational waves may carry evidence 
for the existence of yet unknown fundamental fields that mediate the gravitational interaction together with the spacetime metric as well
as for the existence of new compact objects related to these fundamental fields \cite{Berti:2015itd,Barack:2018yly}. Among the most natural candidates  are the scalar fields that arise naturally in the unified theories in physics. Multiple scalar degrees of freedom are predicted for example by the theories involving extra dimensions. Tensor-multi-scalar theories  (TMST) of gravity are ones of the most viable alternative 
gravitational theories -- they are mathematically self-consistent and can pass all known experimental and observational tests \cite{Damour_1992,Horbatsch_2015}. 
 In general they admit rather rich spectrum of solutions describing various compact objects \cite{Yazadjiev:2019oul,Doneva:2019krb}. 

The main purpose of this paper is to demonstrate that, in the framework of the TMST of gravity, there exists  a particularly interesting
new type of neutron stars. We call this new type of neutron stars {\it topological neutron stars}. In addition to the standard characteristics 
of the usual neutron stars, the topological neutron stars are also characterized by a topological charge. In order  to be specific 
we consider explicit example of a TMST of gravity whose target space is $\mathbb{S}^3$ and numerically construct static and spherically symmetric topological neutron star solutions.

 In  TMST heories the gravitational interaction is mediated  by the spacetime metric $g_{\mu\nu}$ and $N$ scalar fields $\varphi^{a}$  which take value in a coordinate patch of an N-dimensional Riemannian (target) manifold ${\cal E}_{N}$  with positively definite metric $\gamma_{ab}(\varphi)$ defined on it \cite{Damour_1992,Horbatsch_2015}. The Einstein frame action of the  TMST of gravity is given by 
\begin{eqnarray}\label{Action}
S=&& \frac{1}{16\pi G_{*}}\int d^4\sqrt{-g}\left[R - 2g^{\mu\nu}\gamma_{ab}(\varphi)\nabla_{\mu}\varphi^{a}\nabla_{\nu}\varphi^{b} - 4V(\varphi)\right]  \nonumber \\
&&+ S_{matter}(A^{2}(\varphi) g_{\mu\nu}, \Psi_{matter}),
\end{eqnarray}
where $G_{*}$ is the bare gravitational constant, $\nabla_{\mu}$ and $R$ are the covariant derivative  and the Ricci scalar curvature with respect to  the Einstein frame metric $g_{\mu\nu}$, and $V(\varphi)\ge 0$ is the potential of the scalar fields. In order for the weak equivalence principle to be satisfied the matter fields, denoted collectively by $\Psi_{matter}$, are coupled only to the physical Jordan frame metric ${\tilde g}_{\mu\nu}= A^2(\varphi) g_{\mu\nu}$ where  $A^2(\varphi)$ is the conformal factor relating the Einstein and the Jordan frame metrics, and which, together with $\gamma_{ab}(\varphi)$ and $V(\varphi)$, specifies the TMST.  

The Einstein frame field equations  corresponding to the action (\ref{Action}) are the following  
\begin{eqnarray}\label{FE}
&&R_{\mu\nu}= 2\gamma_{ab}(\varphi) \nabla_{\mu}\varphi^a\nabla_{\nu}\varphi^b + 2V(\varphi)g_{\mu\nu} + 8\pi G_{*} \left(T_{\mu\nu} - \frac{1}{2}T g_{\mu\nu}\right), \nonumber \\
&&\nabla_{\mu}\nabla^{\mu}\varphi^a = - \gamma^{a}_{\, bc}(\varphi)g^{\mu\nu}\nabla_{\mu}\varphi^b\nabla_{\nu}\varphi^c 
+ \gamma^{ab}(\varphi) \frac{\partial V(\varphi)}{\partial\varphi^{b}}  \\
&&\hskip 1.6cm -  4\pi G_{*}\gamma^{ab}(\varphi)\frac{\partial\ln A(\varphi)}{\partial\varphi^{b}}T, \nonumber
\end{eqnarray}
with $T_{\mu\nu}$ being  the Einstein frame energy-momentum tensor of matter and $\gamma^{a}_{\, bc}(\varphi)$ being 
the Christoffel symbols with respect to the target space metric $\gamma_{ab}(\varphi)$. From the field equations and the contracted Bianchi identities we also find the following conservation law for the Einstein frame energy-momentum tensor
\begin{eqnarray}\label{Bianchi}
\nabla_{\mu}T^{\mu}_{\nu}= \frac{\partial \ln A(\varphi)}{\partial \varphi^{a}}T\nabla_{\nu}\varphi^a .
\end{eqnarray}

The Einstein frame energy-momentum tensor $T_{\mu\nu}$ and the Jordan frame one ${\tilde T}_{\mu\nu}$
are related via the formula $T_{\mu\nu}=A^{2}(\varphi){\tilde T}_{\mu\nu}$. As usual, in the present paper  the matter content of 
the stars will be described as a perfect fluid. In the case of a perfect fluid the relations between the energy density,
pressure and 4-velocity in both frames are given by $\varepsilon=A^{4}(\varphi){\tilde \varepsilon}$, $p=A^{4}(\varphi){\tilde p}$ and 
$u_{\mu}=A^{-1}(\varphi) {\tilde u}_{\mu}$.

\section{Basic equations and setting the problem}
We are interested in strictly static (in both the Einstein and the Jordan frames), completely regular, spherically symmetric  and asymptotically flat solutions to the equations of the TMST describing neutron stars.  The  spacial slices $\Sigma$
orthogonal to the timelike Killing vector field $\frac{\partial}{\partial t}$ are diffeomorfic to $\mathbb{R}^3$ and the spacetime metric can be written in  the standard form
\begin{eqnarray}
ds^2= - e^{2\Gamma}dt^2 + e^{2\Lambda}dr^2 + r^2(d\theta^2  + \sin^2\theta d\phi^2)
\end{eqnarray} 
where $\Gamma$ and $\Lambda$ depend on the radial coordinate $r$ only.

In the present paper we shall consider TMST whose target space manifold is the round three-dimensional sphere
$\mathbb{S}^3$ with the metric 
\begin{eqnarray}
\gamma_{ab}d\varphi^a d\varphi^b= a^2\left[d\chi^2 + \sin^2\chi(d\varTheta^2 + \sin^2\varTheta d\Phi^2) \right] 
\end{eqnarray}
where $a>0$ is the radius of $\mathbb{S}^3$ and $\chi$, $\varTheta$ and $\Phi$ are the standard angular coordinate on $\mathbb{S}^3$. In addition we shall consider theories for which the coupling function $A(\varphi)$ and the potential $V(\varphi)$ depend on $\chi$ only. Our choice of 
the round $\mathbb{S}^3$ is motivated by the fact that the round $\mathbb{S}^3$ is among the simplest target spaces admitting spherically symmetric topological neutron star solutions.     

We assume that the field $\chi$ depends on the radial coordinate $r$, i.e. $\chi=\chi(r)$, and the fields $\varTheta$ and $\Phi$ are independent from $r$, i.e. $\varTheta=\varTheta(\theta,\phi)$ and $\Phi=\Phi(\theta,\phi)$.  Under these assumptions the equations for $\varTheta$ and $\Phi$ decouple from the rest of the equations and can be considered  separately. Unfortunately the subsystem for $\varTheta$ and $\Phi$ is still rather complicated to be solved analytically and that is why we need one more simplifying assumption, namely $\varTheta=\varTheta(\theta)$
and $\Phi=\Phi(\phi)$. In this case the subsystem for $\varTheta$ and $\Phi$ can be solved in closed analytical form and the solution which is
well-behaved is given by
\begin{eqnarray}\label{ANSATZ}
\varTheta=2 \arctan\left[C\tan^s\frac{\theta}{2}\right], \,\,\,\Phi= s\phi, 	
\end{eqnarray}       
where $s$ is integer and $C>0$ is a constant. However, only the case when $C=1$ and $s=1$ (i.e. $\Theta=\theta$ and $\Phi=\phi$) is compatible with the spherical symmetry and that is why we shall restrict ourselves on it. Then the dimensionally reduced equations are the following
\begin{widetext}
\begin{eqnarray} 
&&\frac{2}{r}e^{-2\Lambda} \Lambda^{\prime} + \frac{1}{r^2}\left(1-e^{-2\Lambda}\right)=8\pi G_{*} A^{4}(\chi) {\tilde \varepsilon}
+a^2 \left(e^{-2\Lambda} {\chi^{\prime}}^2 + 2 \frac{\sin^2\chi}{r^2}\right)  + 2V(\chi), \label{DRE} \\ 
&&\frac{2}{r}e^{-2\Lambda} \Gamma^{\prime} - \frac{1}{r^2}\left(1-e^{-2\Lambda}\right)=8\pi G_{*} A^{4}(\chi) {\tilde p}
+a^2 \left(e^{-2\Lambda} {\chi^{\prime}}^2 - 2 \frac{\sin^2\chi}{r^2}\right)  - 2V(\chi), \\
&&\chi^{\prime\prime} + \left(\Gamma^\prime - \Lambda^\prime + \frac{2}{r}\right)\chi^{\prime}= \left[2\frac{\sin\chi\cos\chi}{r^2}  + \frac{1}{a^2} \frac{\partial V(\chi)}{\partial\chi} + \frac{4\pi G_{*}}{a^2} A^4(\chi)\frac{\partial \ln A(\chi)}{\partial\chi}(\tilde{\varepsilon} - 3{\tilde p})\right]e^{2\Lambda},\\
&& {\tilde p}^\prime = - (\tilde{\varepsilon} + {\tilde p}) \left[\Gamma^\prime + \frac{\partial \ln A(\chi)}{\partial\chi} \chi^\prime \right] \label{HSE}
\end{eqnarray} 
\end{widetext}
where the prime denotes differentiation with repsect to $r$.

Asymptotic flatness requires $\Gamma(\infty)=\Lambda(\infty)=0$ and $\chi(\infty)=k\pi$ with $k$ being integer ($k \in \mathbb{Z}$). Without loss of generality we shall put $k=0$.  Regularity at the center requires $\Lambda(0)=0$ and $\chi(0)=n\pi$ where $n\in \mathbb{Z}$. 
The  integer  $n$ has a topological origin and this can be seen as follows. Since $\chi(\infty)=0$ the map $\varphi :\Sigma\to \mathbb{S}^3$
can naturally be extended to  a map $\varphi :\Sigma\cup\infty \to \mathbb{S}^3$. Taking into account that    $\Sigma\cup \infty = \mathbb{R}^3\cup\infty$ is topologically $\mathbb{S}^3$ we have an effective map $\varphi : \mathbb{S}^3\to \mathbb{S}^3$. The integer $n$ is just the degree of the map $\varphi : \mathbb{S}^3\to \mathbb{S}^3$. Indeed the degree $\deg\varphi$ of the map  $\varphi : \mathbb{S}^3\to \mathbb{S}^3$ is defined as \cite{Bott1982}
\begin{eqnarray}
\deg\varphi=\int_{\Sigma}\varphi_*Vol_{\mathbb{S}^3}
\end{eqnarray}
where $\varphi_*Vol_{\mathbb{S}^3}$ is a 3-form on $\Sigma$ which is the  pull-back  of the normalized volume form  $Vol_{\mathbb{S}^3}$ on the target space $\mathbb{S}^3$ (i.e. $\int_{\mathbb{S}^3} Vol_{\mathbb{S}^3}=1$). A direct calculation shows that $\deg\varphi=n$.

The system of equations (\ref{DRE})--(\ref{HSE}) supplemented with the equation of state of the baryonic matter ${\tilde p}={\tilde p}({\tilde \varepsilon})$, with the above mentioned asymptotic and regularity conditions as  well as with a specified central energy density ${\tilde \varepsilon}$, describes the structure of the neutron stars in the TMST under consideration.

\section{Topological neutron stars}

We shall  consider TMST with $V(\chi)=0$. In this case the asymptotic behavior of the scalar field $\chi$ is given by 
\begin{eqnarray}
\chi\approx \frac{const}{r^2} + O(1/r^3)
\end{eqnarray}
and can be derived from the linearized  equation for $\chi$ outside the star. Let us note that this is nonstandard  asymptotic behavior -- usually the scalar field drops off like $1/r$ far from the stars in the massless scalar-tensor theories. This means that the scalar charge 
$D_{\chi}$ of the neutron star associated with $\chi$ is zero.

All of the neutron star solutions below will be calculated for a realistic nuclear matter equation of state, namely the APR4 EOS \cite{APR4} and we employ its piecewise polytropic approximation \cite{PPA_Read}.

\subsection{Topologically trivial sector $n=0$} 

The topologically trivial sector with $n=0$  possesses a very rich spectrum of physically interesting neutron star solutions in dependence of the coupling function $A(\varphi)$ and the curvature of the target space, i.e. the parameter $a$. Especially interesting is the fact that for certain TMST we can observe spontaneous scalarization. Since the focus of this paper is on the topological neutron stars we shall report our results for the topologically trivial neutron stars in another paper \cite{DY_2019}.

\subsection{Neutron stars with topological charge $n=1$}

Let us first specify the scalar-tensor theory. As an illustrative example we shall focus on the theory with $A(\chi)=\exp(\beta \sin^2\chi)$, where $\beta$ is a (real) parameter.  A numerical solution of the reduced field equations (\ref{DRE})--(\ref{HSE}) together with the above discussed boundary conditions, is obtained via a shooting method. In this subsection we will concentrate on $n=1$, therefore $\chi(r=0)=\pi$ and the central value of the derivative of the scalar field $(d\chi/dr)_0$ is the shooing parameter.

Finding solutions of the field equations is actually a difficult task because we have three free parameters, namely the radius $a$ of the three-dimensional sphere $S^3$, the parameter $\beta$ in the conformal factor $A(\varphi)$ and the central energy density ${\tilde \varepsilon}_c$ of the neutron star. Solutions exist only for certain regions of the parameters space and a thorough examination is required. It is important to note, that more than one solution can exist, i.e. in certain regions of the parameter space we have nonuniqueness. For various combinations of $\beta$ and $a$, we have performed a two dimensional search of solutions with input parameters ${\tilde \varepsilon}_c$ and the shooting parameter $(d\chi/dr)_0$ which guarantees that we have been able to obtain the full spectrum of solutions.

\begin{figure}
	\includegraphics[width=0.45\textwidth]{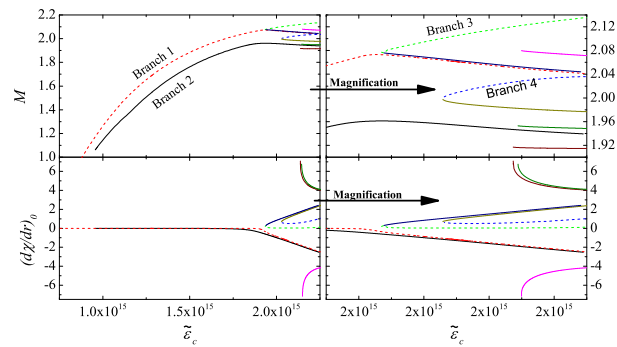}
	\caption{(top panels) The mass as a function of the central energy density for topological charge $n=1$, $a^2=10^{-3}$ and $\beta=0.08$.  The top-right panel is a magnification of the left one around the area where new branches of solutions start to appear. (bottom panels) The shooting parameter $(d\chi/dr)_0$ as a function of the central energy density for the same branches of solutions as in the top panel.  The bottom-right panel is a magnification of the left one.}
	\label{fig:beta0.08_PI}
\end{figure}

It turns out that topological neutron stars can be constructed only for small values of $a$ and in our calculations we have fixed $a^2=10^{-3}$. For a given $a$, there is a minimum $\beta$ below which no topological solutions are present. For small $\beta$ the branches of solutions are short and do not reach a maximum of the mass while for large $\beta$ they can reach such a maximum indicating a change of stability of the corresponding branch. Our studies show that in case such a maximum is reached, new additional branches of solutions appear and the spectrum of solutions becomes very rich. This can be observed in Fig. \ref{fig:beta0.08_PI} (top panels) where the mass and the shooting parameter  $(d\chi/dr)_0$ (bottom panels) are plotted as functions of the central energy density for $\beta=0.08$. The right panels are magnifications of the left ones. As one can see for small ${\tilde \varepsilon}_c$ two branches of solutions are present  but this changes with the increase of the central energy density -- once a maximum of the mass is reached new branches of solutions appear and the number of these additional branches can vary greatly for different  ${\tilde \varepsilon}_c$. We have shown only the branches until roughly ${\tilde \varepsilon}_c=2.2\times 10^{15} {\rm g/cm^3}$ since the structure of solutions becomes more and more rich and complicated for larger ${\tilde \varepsilon}_c$. On the other hand there are some hints that these new branches might be unstable. 

If one compares the two magnified panels in Fig. \ref{fig:beta0.08_PI} depicting the mass and $(d\chi/dr)_0$ as functions of the central energy density, it is evident that  while some of the additional branches merge together with the decrease of  ${\tilde \varepsilon}_c$, others just start from some  ${\tilde \varepsilon}_c$  which corresponds to a divergences in  $(d\chi/dr)_0$. In the figure some of the branches are named with Branch 1-4 and these notations will be used below where particular representative solutions are discussed.

\begin{figure}
	\includegraphics[width=0.35\textwidth]{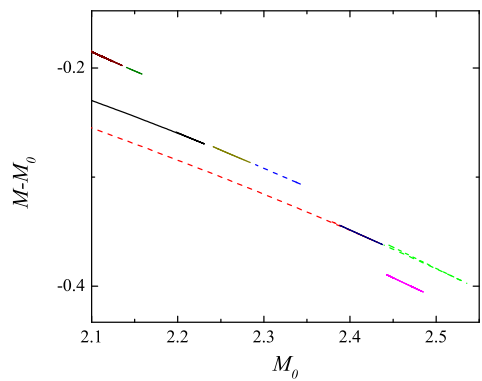}
	\caption{The binding energy $M-M_0$ as a function of the rest mass of the neutron star for topological charge $n=1$, $a^2=10^{-3}$ and $\beta=0.08$.	}
	\label{fig:BindingEn_M0_beta0.08_PI}
\end{figure}

The question which is the more stable branch of solutions can be answered after examining the binging energy $M-M_0$ as a function of the rest mass $M_0$ that is plotted in Fig. \ref{fig:BindingEn_M0_beta0.08_PI}. The branch that has the larger absolute value of the binding energy before a maximum of the mass is reached is the more massive one Branch 1, depicted with red dashed line. It can not be seen clearly in the figure due to the resolution, but all of the branches that reach a maximum mass form a cusp at this maximum-mass point which signals an onset of instability.

A natural question to ask is whether the new additional branches of solutions are stable or not and the rigorous answer can be given after examining the linear stability analysis that will be studied in a subsequent paper. However, our preliminary studies show that Branch1 is stable against linear perturbations up to the maximum of the mass, while the rest of the branches possess unstable modes. Below we will demonstrate that these additional branches have peculiarities in the behavior of the scalar field that influences the rest of the quantities such as the radial profile of the energy density. 

\begin{figure}
	\includegraphics[width=0.4\textwidth]{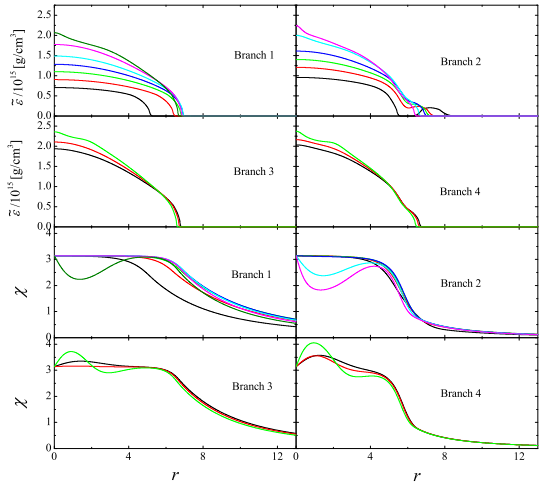}
	\caption{The radial profiles of the scalar field $\chi$ and the energy density for the several representative neutron star models belonging to the different branches in Fig. \ref{fig:beta0.08_PI}, for $n=1$, $a^2=10^{-3}$ and $\beta=0.08$. The numbering of the branches follows the notations introduced in Fig. \ref{fig:beta0.08_PI}. The models that show non-monotonic behavior of the scalar field are typically the ones beyond the maximum of the mass or models belonging to the additional branches. For the potentially stable branch with the largest binding energy, such peculiar behavior happens only after a maximum of the mass is reached.}
	\label{fig:solutions_beta0.08_PI}
\end{figure}

In Fig. \ref{fig:solutions_beta0.08_PI} the radial profiles of the scalar field $\chi$ and the central energy density ${\tilde \varepsilon}$ are shown for neutron stars with different ${\tilde \varepsilon}_c$ belonging to the different branches in Fig. \ref{fig:beta0.08_PI}. The most important feature one can notice is that the solutions belonging to the energetically favorable  Branch 1 have monotonic profiles of $\chi$ and ${\tilde \varepsilon}$ at least for models before the turning point of the mass. The behavior of $\chi$ might not be monotonic any more for larger masses beyond the maximum mass and for the rest of the branches. This influences also the density profiles of the corresponding neutron star solutions which can even have extremum of ${\tilde \varepsilon}(r)$ in the neutron star core. Some of the solutions belonging to the additional energetically less favorable branches exhibit as well a strong minimum of the energy density before the surface of the star is reached. This is a signal that probably part or all of these additional branches are unstable.

\subsection{Neutron stars with topological charge $n=2$}

\begin{figure}
	\includegraphics[width=0.4\textwidth]{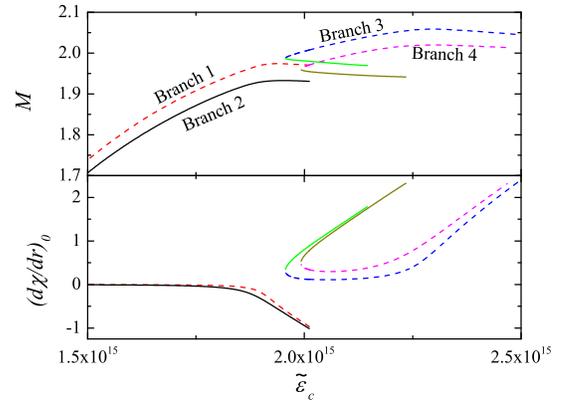}
	\caption{(top panels) The mass as a function of the central energy density for topological charge $n=2$, $a^2=10^{-3}$ and $\beta=0.08$.  The top-right panel is a magnification of the left one around the area where new branches of solutions start to appear. (bottom panels) The shooting parameter $(d\chi/dr)_0$ as a function of the central energy density for the same branches of solutions as in the top panel.  The bottom-right panel is a magnification of the left one.}
	\label{fig:beta0.08_2PI}
\end{figure}

\begin{figure}
	\includegraphics[width=0.35\textwidth]{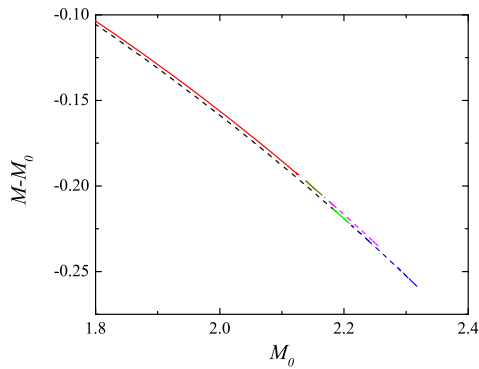}
	\caption{The binding energy $M-M_0$ as a function of the rest mass of the neutron star for topological charge $n=2$, $a^2=10^{-3}$ and $\beta=0.08$.}
	\label{fig:BindingEn_M0_beta0.08_2PI}
\end{figure}

In this subsection we will consider the case when $n=2$ leading to boundary condition for the scalar field $\chi(r=0)=2\pi$. Neutron star solutions with $n=2$ are in general present only for larger values of $\beta$ compared to the $n=1$ case. As a representative example we plot the mass and the shooting parameter for $\beta=0.08$ and $a^2=10^{-3}$ in Fig. \ref{fig:beta0.08_2PI}. One can observe the same qualitative feature as in the $n=1$ case -- two branches of solutions exist for small ${\tilde \varepsilon}_c$ and new solutions appear with the increase of the central energy density beyond the maximum mass points of the first two branches. The main difference with the $n=1$ case is that we have a smaller number of branches for $n=2$ for the same values of $\beta$ and $a$ but this might vary with the change of $\beta$ and $a$. The binding energy for all of these branches is plotted in Fig. \ref{fig:BindingEn_M0_beta0.08_2PI} and again the more stable one is the more massive Branch 1 similar to the $n=1$ case. This seems to be a generic feature for all of the studied combinations of parameters. 
   
\begin{figure}
	\includegraphics[width=0.4\textwidth]{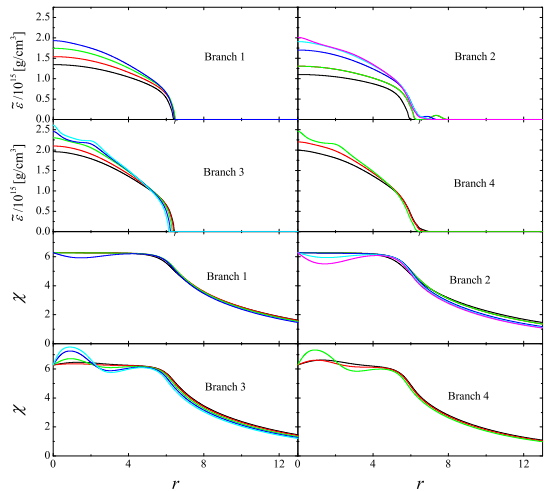}
	\caption{The radial profiles of the scalar field $\chi$ and the central energy density for several representative neutron star models belonging to the different branches depicted in Fig. \ref{fig:beta0.08_PI}, for $n=2$, $a^2=10^{-3}$ and $\beta=0.08$. The central energy density of the chosen models start from relatively small values and reaching beyond the maximum of the mass. We use the numbering of the branches introduced in Fig. \ref{fig:beta0.08_2PI}.   The models that show non-monotonic behavior of the scalar field are typically the ones beyond the maximum of the mass or models belonging to the additional branches.  For the potentially stable branch with the largest binding energy, such peculiar behavior happens only after a maximum of the mass is reached.}
	\label{fig:solutions_beta0.08_2PI}
\end{figure}

 The radial profile of some representative solutions is given in Fig. \ref{fig:solutions_beta0.08_2PI} where the numbering of the branches is the same as in Fig. \ref{fig:beta0.08_2PI}. Similar to the $n=1$ case, the Branch 1 solutions having central energy densities below the maximum mass point  do not show any peculiarities and the energy density and $\chi$ are monotonic functions of $r$. Our preliminary studies show 
 that up to the maximum of the mass Branch1 is linearly stable. This changes for models beyond the maximum mass or for the models belonging to the additional branches where extrema of $\chi$ and even ${\tilde \varepsilon}$ can be present. The appearance of a deep minimum of ${\tilde \varepsilon}$ close to the stellar surface can be observed as well for some of the Branch 2 solutions. All these peculiarities are again a signal that some or all of the branches, except for Branch 1, are unstable. 
 
\subsection{Further investigation of solutions}
In this section we will briefly comment on the further numerical tests we have performed in order to prove the existence of solutions for a wider range of parameters as well as other subclasses of TMST. More specifically, we have examined several cases -- introducing a mass of the scalar field, a change of the conformal factor or changing the sign of $\beta$.

We have studied a scalar field potential of the form $V(\chi)~=~\mu_\chi \sin^2{\chi}/2$, where $m_\chi=\mu_{\chi}/a^2$ is the mass of the field. With the increase of $m_\chi$ the solutions are deformed with respect to the $m_\chi=0$  case but they remain qualitatively similar and after certain $m_{\chi\;\rm crit}$ they disappear, where $m_{\chi\;\rm crit}$ depends on the input parameters. As far as a change of the sign of $\beta$ is concerned, we were able to show that a plethora of solutions exist for $\beta=-0.1$ and $a^2=10^{-3}$ and for reasonable ${\tilde \varepsilon}$ their number is considerably bigger compared to the $\beta>0$ cases examined above.

The second conformal factor we focused on is $A(\chi)=e^{\beta \chi^2/2}$ where $\beta$ is a (positive or negative) constant. We have examined different combinations of $\beta$ and $a$ similar to the $A(\chi)=\exp(\beta \sin^2\chi)$ case. The spectrum of solutions changes significantly but still  several branches of solutions are observed for a proper choice of the parameters. As a matter of fact we were able to find solutions only for values of the parameters that are of the same order as the ones studied in the previous subsection.  

All this shows that the spectrum of solutions is very rich depending on the values of the parameters as well as on the conformal factor and the scalar field potential. Obtaining all branches of solutions only for a single combination of the input parameters is a very time consuming procedure and that is why we have given only some representative examples which prove the existence of such solutions and show their basic properties. Thus the problem deserves further study in parallel with the linear stability analysis. 
  
\section{Conclusion}
 
We have shown that  in certain  classes of TMST of gravity  with $S^3$ target space  there exist topological neutron stars - a new type of neutron stars that possess a topological charge in addition to the other standard characteristics. Besides the topological charge, the topological neutron stars  possess also other attractive features which make them interesting from astrophysical point of view and place them among the realistic compact objects that could exist in Nature. For example they possess zero scalar charge and thus  evades the strong binary pulsar constraints on dipole scalar radiation.

Our preliminary investigations show that there should also exist static non-spherically symmetric but axisymmetric topological neutron stars  which can be found by using the ansatz (\ref{ANSATZ}). The present work can be extended  by considering more complicated target spaces ${\cal E}_{N}$  for which the third homotopy group $\pi_{3}({\cal E}_{N})$ is nontrivial. Then in dependence of the structure of   $\pi_{3}({\cal E}_{N})$ we could have various types of topological neutron stars in the TMST of gravity.

\section*{Acknowledgements}
DD acknowledges financial support via an Emmy Noether Research Group funded by the German Research Foundation (DFG) under grant
no. DO 1771/1-1. DD is indebted to the Baden-Wurttemberg Stiftung for the financial support of this research project by the Eliteprogramme for Postdocs.  SY would like to thank the University of Tuebingen for the financial support.  
The partial support by the Bulgarian NSF Grant DCOST 01/6 and the  Networking support by the COST Actions  CA16104 and CA16214 are also gratefully acknowledged.


\end{document}